\documentstyle{article}
\begin{document}
\begin{center}
{\bf \Large
Social Behaviour of Agents: Capital Markets and Their Small Perturbations
}
\end{center}

\begin{center}
Ondrej Hudak, \footnote{\small present e-mail: hudako@mail.pvt.sk}
\end{center}

\begin{center}
Faculty of Finance, Matej Bel University, \\ Cesta na amfiteater 1, SK-974 15 Banska Bystrica, Slovakia,\\
e-mail: ondrej.hudak@umb.sk
\end{center}

\begin{center}
Jana Tothova
\end{center}

\begin{center}
Stierova 23, Kosice, SK-040 23 Slovakia
\end{center}

\newpage

\section*{Abstract}
We study social behaviour of agents on capital markets when these are perturbed by
small perturbations. We use the mean field method. Social behaviour of agents on capital markets is described:
volatility of the market, aversion constant and equilibrium states are discussed. 
Relaxation behaviour of agents on the capital market is studied. 
Equation of motion for the agent average number is of the relaxation type. Development of the group of agents in the states corresponding to minimum of the aim function is either linear either exponentially damped.
There exist characteristic volatility constants $ V_{c3} $ and $ V_{c3} $.  The constant b of verification of information contribution to the aversion constant A and the $ A_{0} $ constant of aversion are distinguishing three types of dependencies of the minimum of the aim function on the expected volatility EV and on the expected returns E. 
Arbitrage trades and group forces lead the group into the equilibrium state. Verification of information intensity influences return back to the equilibrium state. The linear in time damping to the equilibrium state is characterized with the characteristic time $ T_{3}$ and $  T_{6} $, the exponential with a characteristic time $ \tau $.
Their dependence on the expected volatility, on the expected profit and characteristics of agents is discussed.
\newpage

\tableofcontents{}

\newpage

\newpage
\section{Introduction}
The group of agents on the capital market is characterized as a social group,
\cite{1} - \cite{5.1}. We study information which is private \cite{6}, there is present an information asymmetry.
Unverifiability is a special kind of  information problem \cite{6}. Then there exist subgroups of interacting agents 
on the capital market the interaction of which has the following property: they are nonreducibile \cite{5}.
Nonreducibile subgroups form cells. Configuration of interactions between agents in the group forms a macroscopic structure. The statistical equilibrium due to microreversibility is characterized by the maximum of entropy
and by the minimum of energy (costs of information exchange minus a return).
In \cite{5} we used methods of statistical physics to study  social behaviour of agents.
Three observed dependencies of personal radius of an agent on other characteristics enabled us
to characterize the quantities of cells, faces (nonreducibile subgroups), vertices (agents) and bonds (interactions) \cite{2}. There exist constrains in this structure, a fixed number V of agents in
the group present,  a number E of interactions within the group is constant, a number F
of subgroups which are nonreducibile (faces) is constant, and a number C of cells is constant
for a given conditions as concerning the expected return and expected risk. This structure is equivalent to random cellular networks. Such networks and their evolution were studied in \cite{3} and \cite{4}. 
Random cellular networks describe structure in the group of agents in some approximation, this approximation 
is expected to be sufficient to described main characteristics of the group as concerning the exchange of the information. The capital market and social behaviour of agents was studied recently, \cite{5.1}. 
The quantity which characterizes verification of information contributes to aversion of agent with respect to a risk. The mix of investments of an agent in a given cell with an average measure A of risk aversion in the cell is found from extremes of the function FM, the aim function. We have found the minimum of the average  (per cell) function $<FM>$. An absolute minimum corresponds to such a state in which there is an optimal mix. The number of nonreducible subgroups F per cell $<f>$ contributes linearly to aversion constant A, $ (A = A_{0} + (<f> - 4).b)$. The larger quantity which characterizes verification of information, the larger expected profit is and the larger accepted expected risk is. 
We have formulated model describing behaviour of agents on the capital market \cite{5.1}.
A pyramidal (hierarchical) structure appears in the group for $<f> = 4$. In this case the aversion constant is $A = A_{0} $, which corresponds to hierarchical economy systems. Better verification of information leads to expectation of higher returns and thus the acceptable risk is larger. A total averse constant A is increasing with verification of information and market-based economy system is present in larger extend. We have found that there exists co-existence 
of both economy systems in real economy. In our model the market-based system corresponds to a state with $<n> = 6$ and the hierarchy-based to $<n> = 3$. We assumed that in our model the individual investors and agents from financial intermediary firms do not differ.

The aim of this paper is to study social behaviour of agents on capital markets when these are perturbed.
We study small perturbations of the market using mean field method. Social behaviour of agents on capital markets is described firstly. Model and results of its study, which were done recently by the author \cite{5} and \cite{5.1} for this behaviour are shortly discussed for reference purposes.
Volatility of the market, aversion constant and equilibrium states are discussed then in the next section.
Relaxation behaviour of agents on the capital market in the fourth section is studied. 
Equation of motion for the agent average number is of the relaxational type, and is described in the next section.
Development of the group of agents in the states corresponding to minimum of the aim function is linear and exponentially damped for the case when the extreme of the aim function (which is found from the first derivative of the aim function equal to zero) is minimum and when it is maximum. This development is studied in the sixth section. Conclusions are in the last section.

The main results of the paper are the following.
There exist characteristic volatility constants $ V_{c3} $ and $ V_{c3} $.  The constant b of verification of information contribution to the risk constant A and the $ A_{0} $ constant of aversion are distinguishing three types of dependencies of the minimum of the aim function on the expected volatility EV and on the expected return E. These are
$ A_{0} > 4b $, $ 2b < A_{0} < 4b $ and $ A_{0} < 2b $. The minimum in $ 3 < <n> < 6 $. in $ <n> = 3$ and in $ <n> = 6$
is found. We assumed in this paper that profits are positive. The law of one price is violated by arbitrage opportunities. Arbitrage trades and group forces lead the group into the equilibrium state. The other groups interacting with the group of agents on the capital market are not studied here directly. This group consists of nonreducibile subgroups \cite{5.1}. Nonreducibile subgroups are also groups with their properties. Real markets contain many agents acting on them. We will assume that there is a large number of agents in the group of agents on the capital market. 
In the group of agents on the capital market there are formal groups and informal groups allianced in response to the need for social contact concerning the capital market. Thus interactions of agents described by topology are interactions of agents in formal and informal groups. On the capital market such groups as its subgroups are forming. We mentioned nonreducibile subgroups above, other subgroups are cells which belong to formal type groups and cells which belong to informal type groups. Thus there exist different time periods during which such subgroups exist due to different nature of these subgroups. There are different time periods characteristic for the group of agents on the capital market due to different time to perform different tasks. We assume that processes in the group of agents on capital market are of the relaxation type. There are such reasons for the induced 
fluctuation from the equilibrium (or metastable) state as events in economy and on the market. We assume that these events do not change the equilibrium (or metastable) state. Thus the group of agents has social behaviour for which some of the events in economy and on the market lead to a fluctuation around the minimum state, we assume that this fluctuation is a perturbation of the market of the short time nature at some time t = 0. The equation of motion for the agent average number $ <n> $ per nonreducibile subgroup is found, and is of the relaxation type.
We further assume that the average number $ <n> $ of agents per nonreducibile subgroup is perturbed from the equilibrium state $ <n>´ $. The fluctuation from this state is $ \delta <n> $. The perturbation $ \delta <n>(t) $, which has at the time t = 0 value $ \delta <n>_{0} $, is evolving to zero after some time. We study characteristics of this evolution when the extreme of the aim function is maximum and when this extreme is minimum. We then consider two limiting cases, when the aversion constant $ A_{0} $ is near to 2b from below, $ A_{0} \rightarrow 2b $, and when the aversion constant $ A_{0} $ tends to zero, $ A_{0} \rightarrow 0 $. In the first case we find the relaxation constant $ \tau $  to the lowest order in $ (A_{0} - 2b ) $. When the expected volatility EV increases the relaxation constant $ \tau $ in this limit decreases. When the expected volatility EV decreases the relaxation constant $ \tau $ in this limit increases. When the aversion constant $ A_{0} $ tends to 2b, e.i. when it increases in this limit, the relaxation constant $ \tau $ decreases to zero in the fourth order in $ (A_{0} - 2b) $ .  The constant $ (A_{0} - 4b) $ is called in our paper as the contribution to the total aversion constant from the crowd. It increases when 
the aversion constant $ A_{0} $ tends to 2b. Increasing the aversion constant $ A_{0} $ leads to larger acceptance of the crowd of expected risk, and thus of expected return. Verification of information with its decreasing intensity in this limit leads to slower return back to the equilibrium state. The expected return in this limit and in the first term of the expansion in small parameter is not present in the expression for the relaxation constant $ \tau $.
We obtain also the relaxation constant $ \tau $ to the lowest order in $ (A_{0}  $ tending to zero. When the aversion constant $ A_{0} $ tends to 0, e.i. when it decreases, the relaxation constant $ \tau $ remains finite. 
When the expected volatility EV increases the relaxation constant $ \tau $ in this limit increases. The state with the fluctuation at time $ t = 0 $ returns back to the minimum less quickly. When the expected volatility EV decreases the relaxation constant $ \tau $ in this limit decreases. The state with the fluctuation at time $ t = 0 $ returns back to the minimum more quickly. The larger expected return, the smaller the relaxation constant and the return to the equilibrium is again more quickly. The larger expected volatility, the larger  the relaxation constant and the return to the equilibrium is more slowly. The smaller expected return, the larger the relaxation constant and the return to the equilibrium is more slowly. The larger the intensity b of verification of information the smaller the relaxation constant $ \tau $. The state with a fluctuation at time $ t = 0 $ returns back to the equilibrium state more quickly. The smaller the intensity b of verification of information the larger the relaxation constant $ \tau $. The state with fluctuation returns back to the equilibrium state more slowly. Decreasing the aversion constant $ A_{0} $ leads to smaller acceptance of the expected risk, and thus of the expected return. 
The equation of motion leads in the second case of minimum to the linear in time damping to the equilibrium state with the characteristic time $ T_{3}$, and $  T_{6} $.
For $ A_{0} > 2b $ the characteristic constant $ V_{c6} < 0 $ is negative. 
The characteristic time $ T_{6}$  increases with decrease of risk EV, with decrease of profits, and with decrease of the risk constant $ A_{0}$  toward $ 4b $ . This increase is however to a finite value.
For $ A_{0} < 2b $ the characteristic constant $ V_{c6} > 0 $ is positive. 
The characteristic time $ T_{6}$  increases with decrease of risk EV to the $ V_{c6} $ value or with increase of risk EV to the $ V_{c6} $ value. The increase of the characteristic time $ T_{6}$ is now large.
The characteristic constant $ V_{c3} $ is defined as  $ V_{c3} = 2A_{0}E \frac{A_{0} - 2b}{A_{0} - 4b} $.
With decrease of risk EV toward $ V_{c3} $ the characteristic time $ T_{3} $ increases to large values.

\section{Social Behaviour of Agents: Capital Markets}
We know \cite{5}  that personal diameter r increases when F, the measure of verification of information, decreases. We generalized this relation to every cell: assuming that the personal diameter r increases when $<f>$, the average number of nonreducIbile groups per cell, decreases.  In the topological structure $<f>$ is an average number of faces per cell. Thus generalized quantity $<f>$ is also the quantity which characterizes verification of information. Aversion of an agent to risk on capital markets contains \cite{5.1} a contribution proportional to the verification of information. We assume that the number of nonreducible subgroups F per cell, $<f>$, contributes linearly to this aversion constant A:

\begin{equation}
\label{1}
A = A_{0} - 4.b + <f>.b .
\end{equation}

$ A_{0}$ is the risk aversion constant for $<f> = 4$, then $A = A_{0} $. b is a constant which characterizes contribution
of the quantity $<f>$ to the risk aversion, it is an intensity of contribution of verification of information to the aversion constant. Note that the larger quantity $<f>$ the better verification of information, the larger expected return contribution of the investment and thus the larger acceptable risk. The pyramidal (hierarchical) structure appears in the group when $<f> = 4$, then $A = A_{0} $, \cite{5.1}.
Such a type of structure  corresponds to hierarchical economy systems \cite{6}. Due to very low risk price signals do not work in these systems. Information is flowing through the hierarchy to the decision makers in such structure, however we do not consider the direction of flow of interaction between agents in \cite{5} and \cite{5.1}.
When the structure contains topologically only one cell then $<n> = 6$ and A tends to infinity. Better verification of information leads to expectation of higher returns and thus the acceptable expected risk is larger.
Such a type of structure  corresponds to market-based economy systems \cite{6}. Price signals are essential in these systems. We interpreted $(<f>.b)$ as an average measure of risk aversion of a personal (individual) contribution to the risk aversion. It is different from the crowd contribution of a person in the group, because $<f>$ is a characteristic quantity of verification of information. Agents supply or demand goods and services (on the capital market shares or bonds) if the market price of goods and services exceeds or undervalues his/her own valuation of goods and services (shares or bonds). To make own valuation of shares or bonds the agent individually verifies information about them.
In practice, \cite{6}, both types of economy system involve interaction of people: in the market system they interact  as traders, in the hierarchy system as agents within a hierarchy organization. Both market- and hierarchy- based systems co-exist in modern economy \cite{6}.

In market economies intermediation process is understood quite well in a hypothetical market economy by considering how it might function without financial intermediaries \cite{7}. In our model of capital markets \cite{5.1} every point (vertex) corresponds to an agent acting on the market. In the simplified model economy described in \cite{7}
these agents would be households. Financial intermediaries which lead to lower costs were described in our model 
\cite{5.1} as cell with a given risk aversion constant of the firm to which risk aversion constants of agents from this firm - cell should not be too far. Some other cells are cells with zero risk tolerance constant, agents in these cells correspond to individual savers - agents. We assume that every agent has its objective to maximize its expected
utility of wealth \cite{8}. If returns are normally distributed and the investor has constant absolute risk aversion, then expected utility can be written in an exponential form \cite{9}. Let us denote by A the aversion constant
which when increasing will prefer E, the expected value of end-of-period wealth. EV, the expected variance of end-of-period wealth has then increasing acceptable value. The expected utility of an agent is found as large as possible by the maximization procedure, \cite{10}, assuming that the risk aversion is constant.
In our model \cite{5.1} we assume that every agent has the same E (the expected value of end-of-period wealth) and the same EV (the expected variance of end-of-period wealth). We considered in \cite{5.1} not only maximization of expected utility of an agent, but also optimization of interactions of agents in such a way in which interacting agents maximize their expected utility.
Optimization of the corresponding aim function for given E and EV leads to optimization of the number of agents in the cell and to optimization of verification of information which is characterized by $ <f> $. Then structure of exchange of information in the group of agents will be such that we obtain the highest value of the aim function fM, \cite{5.1}.

The mix of investment of a single cell in the mean field approximation is found from the function:

\begin{equation}
\label{2}
<FM> = (- c.E_{B} + \frac{c^{2}}{2}EV_{B}) \frac{<f>.<n>}{3} + \gamma_{B}c_{B}^{2} \frac{<f>.<n>}{3} ,
\end{equation}

where $E_{B} = E + 2 \gamma c_{B}$ is expected return for a given investment period modified by contribution of the return from the cell (firm) financial intermediary, $EV_{B} = EV + 2 \gamma $ is expected risk for a given investment period modified by contribution of the risk from the cell (firm) financial intermediary, E and EV is expected return and risk respectively for a given investment period. Here $c_{B}$ is a risk constant given for the cell. Note that $ c = \frac{1}{A} $. We study the case $\gamma = 0$, which does not distinguish between individual investors (for which $\gamma = 0$ ) and a financial intermediary firm (for which $\gamma > 0$) .

Thus the equation for the aim function FM is found from (\ref{2}):

\begin{equation}
\label{3}
<FM> = (- c.E + \frac{c^{2}}{2}EV)\frac{<f>.<n>}{3},
\end{equation}

here $\frac{<f>.<n>}{3}$ is an average number of agents per cell. From (\ref{1}) and (\ref{3}) we find:

\begin{equation}
\label{4}
<FM> = (- \frac{E}{A_{0} - 4.b + <f>.b} + \frac{EV}{2.(A_{0} - 4.b + <f>.b)^{2}})\frac{<f>.<n>}{3}
\end{equation}

For an equilibrium structure with a given number of cells C, of faces F, of interactions E, and of agents V, see \cite{3} and \cite{4}, the relation holds:

\begin{equation}
\label{5}
<f>= \frac{12}{(6-<n>)},
\end{equation}

We use $A_{1} = A_{0} - 4.b$. From the aim function FM (\ref{4}) we have found number of cell C, of faces F and of interactions E which minimizes it, in the mean field approximation. The number of agents V is given. It means that there is a minimization of FM with respect to $<n>$, number of interaction in an nonreducibile subgroup  e.i. with respect to $<f>$. We have found \cite{5.1} such a structure of agents on the capital market which gives the lowest value of the aim function FM taking into account their interaction exchanging information.

The function FM has extremes. An absolute minimum corresponds to such a state in which there is an optimal mix, there is an equilibrium state of the market. There exists an extreme of FM, $<n>'$, which is given by \cite{5.1}:

\begin{equation}
\label{6}
<n>'= 6.\frac{(EV (A_{0}-2b)- 2E(A_{0}-2b)^{2})}{( EV.A_{0}-2E(A_{0}-4b)(A_{0}-2b))}.
\end{equation}

The extreme may be an absolute minimum for $A_{0} < 2b$. This is the case when the tendency to verify information is stronger with respect to the aversion constant $A_{0} $. The extreme (\ref{6}) is minimum if the inequalities hold:

\begin{equation}
\label{7}
2E(2b - A_{0})> EV > E \frac{A_{0}}{b} (2b - A_{0})
\end{equation}

Here the state with $<n> = 6$ has lower value of the aim function than the state $<n> = 3$.

The extreme is also minimum if the inequality:

\begin{equation}
\label{7.1}
E \frac{A_{0}}{b} (2b - A_{0}) > EV \geq 0,
\end{equation}

holds. Here the state with $<n> = 3$ has now lower value of the aim function than the state $<n> = 6$.

In both these cases the extreme is a state in which some type of hierarchical structures exists. Either locally, the group is not hierarchical as a whole, in the first case, or it is of the hierarchical structure type in the second case. 
The minimum gives the average number of agents per nonreducibie subgroup which is less than 6 and larger than 3, in the first case, and is equal to 3 in the second case.

The extreme (\ref{6}) is also minimum if:

\begin{equation}
\label{7.2}
EV > 2E(2b - A_{0}).
\end{equation}

Here the state with $<n> = 6$ is minimum. In this case the minimum of the aim function with $<n> = 6$ is a state in which there is no type of hierarchical structure present in the group.

The extreme may be an absolute maximum for $A_{0} > 2b$. In this case the state $<n> = 6$ has always lower value of the aim function than the state with $<n> = 3$.

The extreme (\ref{6}) is maximum if, for $A_{0} > 4b$, the inequality holds:

\begin{equation}
\label{8}
EV > 2E \frac{A_{0}(A_{0}-2b)}{A_{0}-4b}
\end{equation}

There exist two states which minimize the aim function FM. The first one is with $<n> = 3$ (hierarchical) structure.

The second one is for $<n>=6$. The value of the aim function FM is such that this state is much more stable than the state above with $<n> = 3$, which is a metastable state. The number $<n> = 6$ means that there is a one cell structure. Efficient capital markets are ideal markets which are of the this  type (strong stability, conservative and aggressive agents are present on the market, return is high) in our model.

The extreme (\ref{6}) is maximum if, for $A_{0} > 4b$, inequalities hold:

\begin{equation}
\label{8.1}
2E \frac{A_{0}(A_{0}-2b)}{A_{0}-4b} > EV > 2E \frac{(A_{0}-4b)(A_{0}-2b)}{A_{0}}
\end{equation}

There exists one state which minimizes function FM for $3 \leq <n> \leq 6$. The state with $<n> = 3$, which was a metastable state, is now not the metastable state. The minimum of the aim function FM is for $<n>=6$.

The extreme (\ref{6}) is maximum if, for $4b > A_{0} > 2b$, the inequality holds:

\begin{equation}
\label{8.2}
EV < 2E \frac{A_{0}(A_{0}-2b)}{A_{0}-4b}
\end{equation}

There exist two states which minimize function FM. The first one is with $<n> = 3$ and the second one is with $<n>=6$.
This later state has the value of the aim function FM such that this state is much more stable than the state corresponding to $<n> = 3$ which is a metastable state.

The extreme (\ref{6}) is maximum if, for $4b > A_{0} > 2b$, the inequality holds:

\begin{equation}
\label{8.3}
EV > 2E \frac{A_{0}(A_{0}-2b)}{A_{0}-4b}
\end{equation}

The state with $<n>=6$ is now a state corresponding to the minimum of the aim function. There is no metastable state.

\section{Volatility of the Market, Aversion Constant and Equilibrium State}
Let us discuss the relations between the volatility of the market, the aversion constant and the equilibrium state.
When the constant b of verification of information contribution to the risk constant A is such that 
$ A_{0} > 4b $, than for the volatility EV of the price return on the capital market higher that the characteristic constant $ V_{c3} $, e.i. for $EV > V_{c3} $, there is an extreme as maximum present for $ <n> $. The state $ <n> = 3 $, has always higher value of the aim function than the state $ <n> = 6 $. The characteristic constant $ V_{c3} $ is defined as $ V_{c3} = \frac{- A_{0} V_{c6}}{- A_{1}} $. Here the characteristic constant $ V_{c6} $ is defined as $ V_{c6} = - 2E (A_{0} - 2b) $. It depends on the expected return E. The characteristic constant $ V_{c3} $ can be rewritten as $ V_{c3} = \frac{2A_{0}E(A_{0} - 2b) }{- A_{1}} =  \frac{2A_{0}E(A_{0} - 2b) }{- (A_{0} - 4b)} $. For $ A_{0} > 4b $ the characteristic constant $ V_{c6} < 0 $. For $ 0 \leq EV \leq V_{c3}$ there is a maximum at $ <n> = 3$. The $ <n> = 6$ state is the minimum state.
 
When the constant b of verification of information contribution to the risk constant A is such that 
$ 2b < A_{0} < 4b $, than for the volatility EV, which is always positive $ EV > 0 $ (only in the economy with fixed prices the expected volatility is zero $ EV´= 0 $ ) the state $ <n> = 6 $ is always the minimum state.
The state $ <n> = 3 $ is a maximum. The characteristic constant $ V_{c3} $ and the characteristic constant $ V_{c6} $ 
are negative in this case.

When the constant b of verification of information contribution to the risk constant A is such that 
$ 0 \leq A_{0} \leq 2b$, then we find that two characteristic constants $ V_{c3} $ and $ V_{c6}$ obey the inequalities
 $ 0 \leq V_{c3} \leq V_{c6}$. We must distinguish then several cases.
The $ 0 \leq V_{c3} \leq V_{c6} \leq EV$ case is such that the state $ <n> = 6 $ is the minimum state. The state $ <n> = 3 $ is the maximum state.
The $ 0 \leq V_{c3} \leq EV \leq V_{c6} $ case is such that the state $ 3 < <n> < 6 $ is the minimum state, it minimizes the aim function. The state $ <n> = 3 $ is the maximum state, and also the state $ <n> = 6 $ is the maximum state. Depending on whether $ EV \leq V_{cm} $ or whether $ V_{cm} \leq EV $  the state $ <n> = 6 $ is higher than the state $ <n> = 3 $, and the state $ <n> = 3 $  would be higher than the state $ <n> = 6 $, respectively. Here $ V_{cm} $ is an constant defined as $ V_{cm} = \frac{A_{0}V_{c6}}{2b} $. 
The $ 0 \leq EV \leq V_{c3} \leq V_{c6} $ case is such that the state  $ <n> = 3 $ is minimum, the state  $ <n> = 6 $ 
is maximum.

We assumed in this paper and in our previous work \cite{5.1} that profits are positive. The case of negative profits does not correspond to a capital markets on which there is proportional relation between the risk and the positive profit. Here we consider longer term relations.

\section{Relaxation Behaviour of Agents on the Capital Market}
Relaxation behaviour is observed in many social groups. Let us consider arbitrage processes for the group of agents on the capital market. Arbitrage is a way to earn guaranteed profits without investing money \cite{6}. The law of one price says that the same good cannot sell at two different prices on the market. Arbitrage opportunities violate this law. If two assets have identical payoffs  and different prices on the market then this fluctuation of the market will not sustained as an equilibrium state \cite{11}. A profit without risk will not be earned for long time: the buyers and sellers will drive the prices together by trading. Group norms create a force toward behavioural regularities \cite{12}. 
Thus this driving price process is a behavioural regularity. Group cohesiveness creates a force toward behavioural uniformity \cite{12}. Thus arbitrage trades and mentioned forces lead the group into the equilibrium state. We will consider the relaxation process, which after perturbing the group of agents on the market for example by an arbitrage opportunity to the market state with fluctuation, will lead to the equilibrium state.

One of the organizational behaviour  characteristics and aims is productivity \cite{12}. Productivity is a performance measure including effectiveness and efficiency. Effectiveness of the organizational behaviour  is achievements of goals. Efficiency of the organizational behaviour is the ratio of effective output to the input required to achieve it. An organization is productive when it achieves its goals and does so by transferring inputs into outputs with the lowest cost. A group of agents on the capital market is a part of an organization. We consider here agents interacting together exchanging information for given conditions on the capital market. The behaviour of agents in the general group is more than the sum of individual actings in his or her own way. The agents in the group behave in an other way than when they are alone. To understand dynamics of behaviour of the group of agents one should know how an individual, an agent in the group, is influenced by the pattern behaviour he/she is expected to exhibit, what the group considers to be acceptable standard of behaviour, and the degree to which group members are attracted to each other. Organizational system behaviour reaches its highest level of sophistication, \cite{12}, when we add formal structure to knowledge of individual and group behaviour. Organization is more that the sum of their member groups. We study here behaviour of the group of agents on the capital market as mentioned above. The other groups interacting with this group are not studied here directly. This group consists of nonreducibile subgroups \cite{5.1}. nonreducibile subgroups are also groups with their properties. A group is in general formed by two or more individuals, interacting and interdependent, who have come together to achieve some goals. The group of agents on the capital market may have two members, a buyer and a seller. This would be a very elementary case of a market. Real markets contain many agents acting on them. We will assume that there is a large number of agents in the group of agents on the capital market. A group may be formal or informal, \cite{12}. Formal groups are defined by the organizational structure. Informal groups are alliances that are neither formally structured nor organizationally determined. These groups are naturally created groups in the environment that appear  in response to the need for social contacts \cite{12}. In the group of agents on the capital market there are formal groups (financial institutions, firms, other institutions) and informal groups allianced in response to the need for social contact concerning the capital market (agents as individual investors, agents as members of financial institutions, agents as members of firms, agents as members of other institutions, all of them forming also informal groups). There exist in general such groups as command, task, interest, or friendship groups. The last type of the group develops because the individual members have some common characteristics. Social aliances, which may extend outside the work situation, can be based on such common characteristics as is interest about development on the capital market, about development of the price of a share, about development of a firm projects, etc. Thus interactions of agents described by topology in our papers are interactions of agents in formal and informal groups. Agents belong to a number of groups in general. Different groups provide different benefits to their members. Agents are joining into groups due to such reasons as needs for security, status, self-esteem, affiliation, power and goal achievement \cite{12}. This is true in general and this is true also for agents on the capital market. There is no standardized way how a group develops. The Five-Stage Model and the Punctuated-Equilibrium Model are some models describing this development, \cite{12}. In the group of agents on the capital market such groups as its subgroups are forming. We mentioned nonreducibile subgroups above, other subgroups are cells which belong to formal type groups and cells which belong to informal type groups. Thus there exist different time periods during which such subgroups exist due different nature of these subgroups. Some of the groups may be more successful some of them less. It depends on several  variables. Major components that determine a group performance and satisfaction are, \cite{12}, external conditions imposed on the group, group member resources, group structure, the group process-performance/satisfaction relationship  is dependent on the task which the group is working on. These group properties and characteristics hold also for subgroups in the group of agents on capital market. Within a group there are processes such as the communication pattern used by its members to exchange information, group decision processes, leader behaviour, power dynamics, conflict interactions etc \cite{12}. The task uncertainty has the property that nonroutine tasks require considerably more processing of information. Tasks with low uncertainty tend to be standardized. The groups performing first type of the task need to interact with other groups to obtain more and better information. The groups performing the tasks of the second type do not need to interact with other groups so much. Performing task means that after some time this task will be fulfilled.

As we can see there are different time periods characteristic for the group of agents on the capital market due to different time to perform different tasks. While there may be present some oscillation type processes of exchange of information between groups, relaxation of this interaction is expected when the task is fulfilled. Exchange of information within a group may be more oscillatory than for the intergroup case, nevertheless when the task is fulfilled, these interactions will also relax to zero. There are many subgroups in the group of agents on the capital market ending their task in different times. We will assume that processes in the group of agents on capital market are of the relaxation type.

\section{Equation of Motion for the Agent Average Number}
The social group of buyers and sellers is such that its state corresponds to the absolute minimum of the aim function.
The social group of buyers and sellers may be in such a state which corresponds to a metastable minimum of the aim function. Due to many reasons the group may shift from the state corresponding to a minimu of the aim function
to a very near state. There are such reasons for the induced 
fluctuation from the equilibrium (or metastable) state as events in economy and on the market. Arbitrage process is of this type. We will assume that these events do not change the equilibrium (or metastable) state. Thus the group of agents has social behaviour for which some of the events in economy and on the market lead to a fluctuation around the minimum state, we assume that this fluctuation is a perturbation of the market of the short time nature at some time t = 0. We would like to know how the social behaviour of agents return back to the equilibrium (or metastable) minimum state. Thus type of events which may lead to change of the minimum state to other minimum state is not considered in this paper. Also large amplitude perturbations which after some time vanishes are not considered here.

The equation of motion for the agent average number $ <n> $ per nonreducibile subgroup is in this case assumed, as we discussed, of the relaxation type:

\begin{equation}
\label{8.11}
\frac{\partial}{\partial t} \delta <n> = - \lambda \frac{\delta <FM>(<n>´ + \delta <n>)}{\delta \delta <n>} 
\end{equation}

Here we assume that the average number $ <n> $ of agents per nonreducibile subgroup is perturbed from the equilibrium state 
$ <n>´ $. The fluctuation from this state is $ \delta <n> $. Time t is in general in the interval $ 0 \leq t \leq \infty $. The perturbation $ \delta <n>(t) $, which has at the time t = 0 value $\delta <n>_{0} $, is evolving to zero after some time. We will study characteristics of this evolution when the extreme of the aim function is maximum and when this extreme is minimum. In the first case we have found the minimum at $ <n>´= 3 $  and at $ <n>´= 6 $. In the second case the minimum is such that $ 3 < <n>´< 6 $, other minimum is at $ <n>´= 3 $, or at $ <n>´= 6 $

\section{Development of the Group of Agents: The Extreme is Minimum}

When the extreme (\ref{6}) is minimum, then the state $ 3 < <n> < 6$  or the state with $<n> = 3$ which has lower value of the aim function than the state $<n> = 6$ or the state with $<n> = 6$ which has lower value of the aim function than the state $<n> = 3$ , are equilibrium states. The equation of motion leads in the first case to the exponential damping of the fluctuation to the equilibrium state with the relaxation time $ \tau $:

\begin{equation}
\label{8.21}
\delta <n> = \delta <n>_{0} \exp(- \frac{t}{\tau})
\end{equation}

here $ \delta <n>_{0} $ is a fluctuation of the state at time zero , and where:

\begin{equation}
\label{8.31}
\tau =  \frac{144 (EV b)^{2}(A_{0} - 2b)} { \lambda (EV.A_{0} - 2E(A_{0} - 2b)(A_{0} - 4b))^{2}}.S
\end{equation}

and where the quantity S is defined as:
 
\begin{equation}
\label{8.211}
S = - (EV + 2E(A_{0} - 4b)) + 
\frac{(A_{0} - 4b)^{2}}{192 EV.b^{2} (A_{0} - 2b)^{2}} [ 96 b (A_{0} - 2b)(- EV^{2} + 2E^{2}(A_{0} -2b)^{2}) 
\end{equation}
$ + (EV.A_{0} - 2E(A_{0} - 2b)(A_{0} - 4b))^{2}] + \frac{E(A_{0}^{2} - (4b)^{2})}{2b} $

Let us consider two limiting cases, when the aversion constant $ A_{0} $, is near to 2b from below, $ A_{0} \rightarrow 2b $, and when the aversion constant $ A_{0} $ tends to zero, $ A_{0} \rightarrow 0 $.

In the first case we obtain that the relaxation constant $ \tau $ is, to the lowest order in $ (A_{0} - 2b ) $, given by:

\begin{equation}
\label{8.311}
 \tau =  \frac{486 b^{2} (A_{0} - 2b)^{4}}{\lambda . EV . A_{0}^{4}}
 \end{equation}
 
When the expected volatility EV increases the relaxation constant $ \tau $ in this limit decreases. The state with the fluctuation at time $ t = 0 $ returns back to the minimum more quickly. 
When the expected volatility EV decreases the relaxation constant $ \tau $ in this limit increases. The state with the fluctuation at time $ t = 0 $ returns back to the minimum less quickly. The expected volatility EV influences how quickly
the state with fluctuation returns back to the equilibrium state.
When the aversion constant $ A_{0} $ tends to 2b, e.i. when it increases in this limit, the relaxation constant $ \tau $
decreases to zero in the fourth order in $ (A_{0} - 2b) $ .  The constant $ (A_{0} - 4b) $ is called in our paper as the contribution to the total aversion constant from the crowd. It increases when 
the aversion constant $ A_{0} $ tends to 2b. The state with fluctuation decreases more quickly to the equilibrium state.
The ability of the crowd to remain self-aware and think logically is suppressed \cite{1}. Increasing the aversion constant $ A_{0} $ leads to larger acceptance of the crowd of expected risk, and thus of expected return. 
When the intensity of personal contribution to the aversion constant, b, decreases in this limit,then the relaxation constant $ \tau $ firstly increases, and then for $ b < \frac{A_{0}}{6} $ decreases. This later inequality is however such that the expansion in small parameter $ A_{0} - 2b \propto \frac{2A_{0}}{3} $ is not valid in general. Thus verification of information with its decreasing intensity in this limit leads slower return back to the equilibrium state.
Note that the expected return in this limit and in the first term of the expansion in small parameter is not present in the expression for the relaxation constant $ \tau $.

In the second case we obtain that the relaxation constant $ \tau $ is, to the lowest order in $ A_{0}  $, given by:

\begin{equation}
\label{8.32}
 \tau =  \frac{27 . EV^{3}}{32 \lambda E^{4} b^{2} }
 \end{equation}
 
When the expected volatility EV increases now the relaxation constant $ \tau $ in this limit increases. The state with the fluctuation at time $ t = 0 $ returns back to the minimum less quickly. 
When the expected volatility EV decreases the relaxation constant $ \tau $ in this limit decreases. The state with the fluctuation at time $ t = 0 $ returns back to the minimum more quickly. The expected volatility EV influences how quickly
the state with fluctuation returns back to the equilibrium state also in this case. 
When the aversion constant $ A_{0} $ tends to 0, e.i. when it decreases, the relaxation constant $ \tau $
remains finite.  The constant $ (A_{0} - 4b) $,the contribution to the total aversion constant from the crowd, becomes 
larger in amplitude and negative. The crowd behaves more irrationally. As we mentioned above the smaller expected volatility, the smaller the relaxation constant and the return to the equilibrium is more quickly. And the larger expected return, the smaller the relaxation constant and the return to the equilibrium is again more quickly. The larger expected volatility, the larger  the relaxation constant and the return to the equilibrium is more slowly. The smaller expected return, the larger the relaxation constant and the return to the equilibrium is more slowly. 
The ability of the crowd to remain self-aware and think logically is suppressed \cite{1} as we mentioned. The dominance of the crowd´s belief system limits severe quality of data which crowd recognizes as information \cite{1}. The larger the intensity b of verification of information the smaller the relaxation constant $ \tau $. The state with a fluctuation at time $ t = 0 $ returns back to the equilibrium state more quickly. The smaller the intensity b of verification of information the larger the relaxation constant $ \tau $. The state with fluctuation returns back to the equilibrium state more slowly.
Decreasing the aversion constant $ A_{0} $ leads to smaller acceptance of the expected risk, and thus of the expected return.

The equation of motion leads in the second case of minimum to the linear in time damping to the equilibrium state with: the characteristic time $ T_{3}$:

\begin{equation}
\label{8.3111}
\delta <n> = \delta <n>_{0}(1 - \frac{t}{T_{3}})
 \end{equation}
 
 where:

\begin{equation}
\label{8.4}
T_{3} = \frac{\mid \delta <n>_{0} \mid 3 A_{0}^{2}}{\lambda \mid (EV - V_{c3})(A_{0} - 4b) \mid}  
 \end{equation}
 
Here $ \delta <n>_{0} $ is again a fluctuation of the state at time zero.

When the state $<n> = 6$ is an equilibrium state, the equation of motion lead in this case to the linear in time damping with the characteristic  time $ T_{6} $:

\begin{equation}
\label{8.41}
\delta <n> = \delta <n>_{0}(1 - \frac{t}{T_{6}})
 \end{equation}
 
 where:

\begin{equation}
\label{8.42}
T_{6} = \frac{\mid \delta <n>_{0} \mid 12 \delta^{2}}{\lambda \mid V_{c6} - EV \mid}
 \end{equation}

Discussion of dependencies of characteristic times $ T_{3} $ and $ T_{6} $ are formally the same as in the case when the extreme is maximum, see below.

\section{Development of the Group of Agents: The Extreme is Maximum}
When the extreme (\ref{6}) is maximum, then the state $ < <n> = 6$  or the state with $<n> = 3$ are the states corresponding to minimum of the aim function.  In this case the state $<n> = 6$ has always lower value of the aim function than the state with $<n> = 3$. Thus the state $<n> = 6$ is an equilibrium state, the state with $<n> = 3$
is a metastable state. As discussed above the state $<n> = 6$ has always lower value of the aim function than the state with $<n> = 3$. In the case discussed above the state $<n> = 6$ is an equilibrium state, and the state with $<n> = 3$
is not a metastable state, it is a maximum.

The equation of motion lead in the first case to the linear in time damping with the characteristic  time $ T_{6}$:

\begin{equation}
\label{8.5}
\delta <n> = \delta <n>_{0}(1 - \frac{t}{T_{6}})
 \end{equation}
 
where:

\begin{equation}
\label{8.6}
T_{6} = \frac{\mid \delta <n>_{0} \mid 12 \delta^{2}}{\lambda \mid V_{c6} - EV \mid}
 \end{equation}
 
Here $ \delta <n>_{0} $ is again a fluctuation of the state at time zero.
For $ A_{0} > 2b $ the characteristic constant $ V_{c6} < 0 $ is negative. 
The characteristic time $ T_{6}$  increases with decrease of risk EV, with decrease of profits, and with decrease of the risk constant $ A_{0}$  toward $ 4b $ . This increase is however to a finite value.
For $ A_{0} < 2b $ the characteristic constant $ V_{c6} > 0 $ is positive. 
The characteristic time $ T_{6}$  increases with decrease of risk EV to the $ V_{c6} $ value or with increase of risk EV to the $ V_{c6} $ value. The increase of the characteristic time $ T_{6}$ is now large.

The equation of motion leads in case of the metastable state case $<n> = 3$ also to the linear in time damping to the locally minimum state with the characteristic time $ T_{3}$:

\begin{equation}
\label{8.7}
\delta <n> = \delta <n>_{0}(1 - \frac{t}{T_{3}})
\end{equation}
 
 where:

\begin{equation}
\label{8.411}
T_{3} = \frac{\mid \delta <n>_{0} \mid 3 A_{0}^{2}}{\lambda \mid (EV - V_{c3})(A_{0} - 4b) \mid}  
\end{equation}
 
Here $ \delta <n>_{0} $ is again a fluctuation of the state at time zero.
The characteristic constant $ V_{c3} = 2A_{0}E \frac{A_{0} - 2b}{A_{0} - 4b} $.
With decrease of risk EV toward $ V_{c3} $ the characteristic time $ T_{3} $ increases to large values,

The state $<n> = 6$ has always lower value of the aim function than the state with $<n> = 3$. In the case discussed above when the state $<n> = 6$ is an equilibrium state, and the state with $<n> = 3$ is not a metastable state, the characteristic time $ T_{6} $ discussion above is the same. There is no linear in time damping to the state $<n> = 3$,
it is a maximum.

\section{Conclusions}
We studied capital market and social behaviour of agents. Quantity F, and in general $<f>$, is the
quantity which characterizes verification of information and thus contributes to aversion of an agent with respect to a risk. We generalized this relation to every cell: we assumed that personal diameter r increases when $<f>$, the average number of nonreducibile groups per cell, decreases, $<f>$ is an average number of faces per cell. Aversion of an agent to risk on capital
markets also contains characterization of verification of information. We assumed that the number of nonreducible subgroups F per cell, $<f>$, contributes linearly to this aversion constant A.

When the pyramidal structure appears in the group, $<f> = 4$, then $A = A_{0} $.
This type of structure  corresponds to hierarchical economy systems. When the structure
contains topologically only one cell then $<n> = 6$ and A tends to infinity. Better verification of information leads to expectation of higher returns and thus the acceptable risk is larger.
This type of structure  corresponds to market-based economy systems. They work via price signals.
We interpret $A_{0}  -  4.b$ as a kind of measure of risk aversion of crowd and we interpret $(<f>.b)$
as an average measure of risk aversion of a personal /individual/ contribution to risk aversion different from the crowd contribution of this person in the cell. 

In practice both types of system involve interaction of people: in the market system they interact  as traders, in the hierarchy as agents within an organization. Both market- and hierarchy- based systems co-exist in modern economy: we have open markets and we also have large organizations.
This coresponds to our state which is minimum for $ 3 < <n> <6$. In our model the market-based system corresponds to a state with $<n> = 6$ and the hierarchy-based to $<n> = 3$.

The mix of investments of an agent in a given cell with an average measure A of risk aversion in the cell is found from
extremes of the function FM, the aim function. We find the minimum of the average  /per cell/ function $<FM>$. We assume that in our model the individual investors and agents from financial intermediary firms do not differ. More general case will studied elsewhere.

When $<f> = 4$ then the pyramidal (hierarchical) structure exists in which there are only nonreducibie subgroups with $n = 3$. Thus we consider this structure as a structure with a uniform risk aversion constant $A_{0}$. Then a
personal contributions to the risk aversion constant of an agent are those contributions which are corresponding
to structures with $<n>  > 3$. Thus there is a linear contribution to the risk aversion constant which is proportional to $<f> - 4$. This corresponds to nonuniform risk aversion constants in the structure. This contains a personal contribution to
the risk aversion constant.
We use $\frac{<f>.<n>}{3}$ as an average number of agents per cell.

The extreme of $<FM>$ is minimum for some conditions on the market, and in the group of agents, see above. This minimum corresponds to $<n>$ between 3 to 6. This is a stable state. The extreme may be a maximum. Then there are two minima of the function FM, one of them is that which corresponds to $<n> = 3$ (the pyramidal hierarchical structure) which is a metastable state, and the other one corresponds to $<n> = 6$  (the structure with hexagons in average and with one cell) which is a stable state. Efficient capital markets are markets which are of the second type (strong stability, conservative and aggressive agents are present on the market, returns are high). Aboav's law describes how many, $m(f,n)$, agents are present in average in nonreducibie subgroups in a cell neighbouring to a cell with f nonredducibile subgroups with agent average number n. This enables us to characterize variance of n for the distribution of nonreducibile subgroups. This variance diverges to infinity with $<n>$ tending to 6.

The financial intermediaries play a key role to reduce the risks faced by individual savers, by pooling their savings and using these to assemble diversified portfolios of assets. This diversification requires specialized knowledge and expertise, and a large portfolio. This will studied using our model with nonzero $\gamma $ constant coresponding to a firm with a given tolerance risk constant.

The aim of this paper was to study social behaviour of agents on capital markets when these are perturbed.
We studied small perturbations of the market using mean field method. Social behaviour of agents on capital markets was described firstly. Model and results of its study, which were done recently by the author \cite{5} and \cite{5.1}, for this behaviour were shortly discussed for reference purposes. Volatility of the market, aversion constant and equilibrium states are discussed then in the next section. Relaxation behaviour of agents on the capital market in the fourth section is studied. Equation of motion for the agent average number is of the relaxational type, and is described in the next section. The oscillation type of motion is not considered here.
Development of the group of agents in the states corresponding to minimum of the aim function is linear and exponentially damped for the case when the the extreme of the aim function is minimum and is maximum. This development 
was studied in the sixth section. 

We conclude that there exist characteristic volatility constants $ V_{c3} $ and $ V_{c3} $.  The constant b of verification of information contribution to the risk constant A and the $ A_{0} $ constant of aversion are distinguishing three types of dependencies of the minimum of the aim function on the expected volatility EV and on the expected returns E. These are
$ A_{0} > 4b $, $ 2b < A_{0} < 4b $ and $ A_{0} < 2b $. The minimum in $ 3 < <n> < 6 $. in $ <n> = 3$ and in $ <n> = 6$
is found. Thus estimating the average values of the aversion constant $ A_{0} $ and of the constant b of the intensity of the verification of information we should distinguish which type of dependence of the equilibrium market state occurs.
The other groups interacting with the group of agents on the capital market are not studied here directly. 
Their inclusion would lead to study of the intergroup interactions. On the capital market subgroups are forming. We mentioned nonreducibile subgroups above, other subgroups are cells which belong to formal type groups and cells which belong to informal type groups. We assume that processes in the group of agents on capital market are of the relaxation type. There are such reasons for the induced fluctuation from the equilibrium (or metastable) state as events in economy and on the market. We assume that these events do not change the equilibrium (or metastable) state. Thus the group of agents has social behaviour for which some of the events in economy and on the market lead to a fluctuation around the minimum state, we assume that this fluctuation is of a perturbation of the market of the short time nature at some time t = 0. The equation of motion for the agent average number $<n> $ per nonreducibile subgroup is found, and is of the relaxation type. We studied characteristics of evolution of perturbation on the market when the extreme of the aim function is maximum and when this extreme is minimum. We consider two limiting cases, when the aversion constant $ A_{0} $, is near to 2b from below, $ A_{0} \rightarrow 2b $, and when the aversion constant $ A_{0} $ tends to zero, $ A_{0} \rightarrow 0 $. In the first case we obtain that the relaxation constant $ \tau $ is, to the lowest order in $ A_{0} - 2b ) $. When the expected volatility EV increases the relaxation constant $ \tau $ in this limit decreases. When the expected volatility EV decreases the relaxation constant $ \tau $ in this limit increases. 
When the aversion constant $ A_{0} $ tends to 2b, e.i. when it increases in this limit, the relaxation constant $ \tau $ decreases to zero in the fourth order in $ (A_{0} - 2b) $ .  The constant $ (A_{0} - 4b) $ is called in our paper as the contribution to the total aversion constant from the crowd. It increases when 
the aversion constant $ A_{0} $ tends to 2b. Increasing the aversion constant $ A_{0} $ leads to larger acceptance of the crowd of expected risk, and thus of expected return. Verification of information with its decreasing intensity in this limit leads slower return back to the equilibrium state. The expected return in this limit and in the first term of the expansion in small parameter is not present in the expression for the relaxation constant $ \tau $.
We obtain the relaxation constant $ \tau $ to the lowest order in $ (A_{0}  $ tending to zero. 
When the expected volatility EV increases now the relaxation constant $ \tau $ in this limit increases. The state with the fluctuation at time $ t = 0 $ returns back to the minimum less quickly. When the expected volatility EV decreases the relaxation constant $ \tau $ in this limit decreases. The state with the fluctuation at time $ t = 0 $ returns back to the minimum more quickly. When the aversion constant $ A_{0} $ tends to 0, e.i. when it decreases, the relaxation constant $ \tau $ remains finite.  The larger expected return, the smaller the relaxation constant and the return to the equilibrium is again more quickly. The larger expected volatility, the larger  the relaxation constant and the return to the equilibrium is more slowly. The smaller expected return, the larger the relaxation constant and the return to the equilibrium is more slowly. The larger the intensity b of verification of information the smaller the relaxation constant $ \tau $. The state with a fluctuation at time $ t = 0 $ returns back to the equilibrium state more quickly. The smaller the intensity b of verification of information the larger the relaxation constant $ \tau $. The state with fluctuation returns back to the equilibrium state more slowly. Decreasing the aversion constant $ A_{0} $ leads to smaller acceptance of the expected risk, and thus of the expected return. 
The equation of motion leads in the second case of minimum to the linear in time damping to the equilibrium state with the characteristic time $ T_{3}$ and $  T_{6} $.
For $ A_{0} > 2b $ the characteristic constant $ V_{c6} < 0 $ is negative. 
The characteristic time $ T_{6}$  increases with decrease of risk EV, with decrease of profits, and with decrease of the risk constant $ A_{0}$  toward $ 4b $ . This increase is however to a finite value.
For $ A_{0} < 2b $ the characteristic constant $ V_{c6} > 0 $ is positive. 
The characteristic time $ T_{6}$  increases with decrease of risk EV to the $ V_{c6} $ value or with increase of risk EV to the $ V_{c6} $ value. The increase of the characteristic time $ T_{6}$ is now large.
The characteristic constant $ V_{c3} = 2 A_{0} E \frac{A_{0} - 2b}{A_{0} - 4b} $.
With decrease of risk EV toward $ V_{c3} $ the characteristic time $ T_{3} $ increases to large values, 
Our model thus, under defined conditions describes development on the market, when small perturbations are present. 
The developed capital markets are different from emerging capital markets, \cite{13}. It is interesting, that we have found and exponentially damped behaviour of the index on the Bratislava Stock Exchange and Prague Stock Exchange, \cite{14}, when the trading started on these markets after coupon privatization ended. The prices of the most liquid shares were in that moment still not in equilibrium, they developed after several months to an equilibrium state. 
It is well known property of prices (of commodities physical and financial) that when after increase of the price due to some perturbation these prices return back to the equilibrium price more slowly than when after decrease of the price of the commodity due to some perturbation these prices return back to the equilibrium price. Thus the relaxation constant of the first return back is larger than of the second one. While the expected price is in both cases the same, we can say within our model, assuming that we are in the limit of low $ A_{0} $, that the expected volatility is lower in the second case than in the first second one. The quantitative and qualitative dependencies described above may be thus compared with those observed on real markets.

\section*{Acknowledgment}

The paper represents a part of results of the VEGA project 1/0495/03.

\end{document}